\documentstyle[preprint,pre,aps,oldlfont]{revtex}
\begin{document}
\draft
\title{
       $M$-furcations in coupled maps
      }
\author{Sang-Yoon Kim\footnote{Electronic address: sykim@cc.kangwon.ac.kr}}
\address{
Department of Physics \\ Kangwon National University\\
Chunchon, Kangwon-Do 200-701, Korea
}
\maketitle

\begin{abstract}
We study the scaling behavior of $M$-furcation $(M\!=\!2, 3, 4,\dots)$
sequences of $M^n$-period $(n=1,2,\dots)$ orbits
in two coupled one-dimensional (1D) maps.
Using a renormalization method, how the scaling behavior
depends on $M$ is particularly investigated
in the zero-coupling case in which the two 1D maps become uncoupled.
The zero-coupling fixed map of the $M$-furcation renormalization
transformation is found to have three relevant eigenvalues $\delta$, $\alpha$,
and $M$ ($\delta$ and $\alpha$ are the parameter and orbital scaling factors
of 1D maps, respectively). Here the second and third ones, $\alpha$ and $M$,
called the ``coupling eigenvalues'', govern the scaling behavior associated
with coupling, while the first one $\delta$ governs the scaling behavior
of the nonlinearity parameter like the case of 1D maps.
The renormalization results are also confirmed by a direct
numerical method.
\end{abstract}
\pacs{PACS numbers: 05.45.+b, 03.20.+i, 05.70.Jk}

%
%

\narrowtext

\section{Introduction}
\label{sec:Int}

Universal scaling behaviors of $M$-furcation $(M=2,3,4,\dots)$ sequences
of $M^n$-cycles $(n=1,2,\dots)$ (i.e., $M^n$-period orbits)
have been found in a one-parameter family of one-dimensional (1D)
unimodal maps with a quadratic maximum. As an example, consider the
logistic map
\begin{equation}
x_{t+1} = f(x_t) = 1 - A\, x_t^2,
\label{eq:1DM}
\end{equation}
where $t$ denotes the discrete time.
As the nonlinearity parameter $A$ is increased from $0$, a stable fixed
point undergoes the cascade of period-doubling bifurcations accumulating
at a finite parameter value $A_{\infty}(=1.401\,155\, \dots)$.
The bifurcation sequence corresponding to
the MSS (Metropolis, Stein, and Stein \cite{Metropolis}) sequence
$R^{*n}$
(for details of the MSS sequences and the $(*)$-composition rule, see
Refs.~\cite{Metropolis,Derrida}) exhibits an asymptotic scaling behavior
\cite{Feigenbaum}.

What happens beyond the bifurcation accumulation point $A_\infty$ is
interesting from the viewpoint of chaos.
The parameter interval between $A_{\infty}$ and the final boundary-crisis
point $A_c(=2)$ beyond which no periodic or chaotic attractors can be found
within the unimodality interval is called the ``chaotic'' regime. Within this
region, besides the bifurcation sequence, there are many other sequences of
periodic orbits exhibiting their own scaling behaviors. In particular, every
primary pattern $P$ (that cannot be decomposed using the ($*$)-operation)
leads to an MSS sequence $P^{*n}$.
For example, $P=RL$ leads to a trifurcation sequence of $3^n$-cycles,
$P=RL^2$ to a tetrafurcation sequence of $4^n$-cycles,
and the three different $ P= RLR^2,\;RL^2R$, and $RL^3$ to three different
period-$5^n$ sequences. Thus there exist infinitely
many higher $M$-furcation $(M=3,4,\dots)$ sequences inside the chaotic regime.
Unlike the bifurcation sequence, stability regions of periodic orbits
in the higher $M$-furcation sequences are not adjacent on the parameter
axis, because they are born by their own tangent bifurcations. The
asymptotic scaling behaviors of these (disconnected) higher $M$-furcation
sequences characterized by the parameter and orbital scaling factors,
$\delta$ and $\alpha$, vary depending on the primary pattern $P$
\cite{Derrida,Hu,Eckmann,Hao,Hauser,Chang,Kenny1,Kenny2,Urumov}.

In this paper we consider two symmetrically coupled 1D maps. This
coupled map may help us to understand how coupled oscillators,
such as Josephson-junction arrays or chemically reacting cells, exhibit
various dynamical behaviors \cite{Kaneko,Yuan,Hogg}. We are interested in
the scaling behaviors
of $M$-furcations $(M=2,3,\dots)$ in the two coupled 1D maps.
The bifurcation case $(M=2)$ was previously studied in
Refs.~\cite{Kuznetsov,Kook,Kim1,Kim2,Kim3,Kim4}.
Here we extend the results for the bifurcation case to all the other higher
multifurcation cases with $M=3,4,\dots$ in the
zero-coupling case where the two 1D maps become uncoupled.
In Sec.~\ref{sec:RA} we investigate the dependence of the scaling behavior
on $M$ using a renormalization method.
It is found that the zero-coupling
fixed point of the $M$-furcation renormalization transformation has three
relevant eigenvalues $\delta$, $\alpha$, and $M$.
The scaling behavior associated with coupling is governed by two coupling
eigenvalues (CE's) $\alpha$ and $M$, while the scaling behavior of the
nonlinearity parameter is also governed by the eigenvalue $\delta$ like
the case of 1D maps. As an example, we numerically study the scaling
behavior associated with coupling in the trifurcation sequence in
Sec.~\ref{sec:NA} and confirm
the renormalization results. Finally, a summary is given in
Sec.~\ref{sec:summary}.

\section {Renormalization analysis}
\label{sec:RA}

In this section we first introduce two coupled 1D maps and discuss stability
of orbits, and then study the scaling behavior of $M$-furcations
$(M=2,3,\dots)$ in the zero-coupling case using the renormalization method
developed in Refs.~\cite{Kuznetsov,Kim3}.
It is found that there exist three relevant eigenvalues
$\delta$, $\alpha$, and $M$. As in the case of 1D maps, the scaling behavior
of the nonlinearity parameter is governed by the eigenvalue $\delta$,
irrepectively of coupling. However, the scaling behavior associated with
coupling depends on the nature of coupling. In a linear-coupling case,
in which the coupling function has a leading linear term, it is governed
by two CE's $\alpha$ and $M$, whereas it is governed by only one CE $M$ in
the other cases of nonlinear coupling with leading nonlinear terms.

Consider a map $T$ consisting of two symmetrically coupled 1D maps,
\begin{equation}
   T:\left\{
       \begin{array}{l}
        x_{t+1} = F(x_t,y_t) = f(x_t) + g(x_t,y_t), \\
        y_{t+1} = F(y_t,x_t) = f(y_t) + g(y_t,x_t),
       \end{array}
     \right.
\label{eq:CM}
\end{equation}
where $f(x)$ is a 1D unimodal map with a quadratic maximum at $x=0$, and
$g(x,y)$ is a coupling function. The uncoupled 1D map $f$
satisfies a normalization condition
\begin{equation}
f(0) = 1,
\label{eq:NC}
\end{equation}
and the coupling function $g$ obeys a condition
\begin{equation}
g(x,x) = 0 \;\;\;{\rm for \; any \;} x.
\label{eq:CC}
\end{equation}

The two-coupled map (\ref{eq:CM}) is invariant under the exchange of
coordinates such that $x \leftrightarrow y$. The set of all points which
are invariant under the exchange of coordinates forms a symmetry line
$y=x$. An orbit is called an ``in-phase'' orbit if it lies on the
symmetry line, i.e., it satisfies
\begin{equation}
x_t = y_t\;\;\;{\rm for \; all\;} t.
\label{eq:IO}
\end{equation}
Otherwise it is called an ``out-of-phase'' orbit. Here we study only
in-phase orbits, which can be easily found from the uncoupled 1D map,
$x_{t+1} = f(x_t)$, because of the condition (\ref{eq:CC}).

Stability of an in-phase orbit with period $p$ is
determined from the Jacobian matrix $J$ of $T^p$,
which is the $p$-product of the Jacobian matrix $DT$ of $T$
along the orbit:
\begin{equation}
 J = {\prod_{t=1}^{p}} DT(x_t,x_t)= {\prod_{t=1}^{p}}
                                   \left( \begin{array}{cc}
                                   f'(x_t) - G(x_t) & G(x_t)\\
                                   G(x_t) & f'(x_t)-G(x_t)
                                   \end{array}
                                    \right),
\label{JAC}
\end{equation}
where the prime denotes a derivative, and
$G(x) = {\partial g}(x,y)/{\partial y} \mid_{y=x}$;
hereafter, $G(x)$ will be referred to as the ``reduced
coupling function'' of $g(x,y)$.
The eigenvalues of $J$, called the stability multipliers of the orbit, are:
\begin{equation}
\lambda_1 = \prod_{t=1}^p f'(x_t), \;\;\;
\lambda_2 = \prod_{t=1}^p [f'(x_t)-2G(x_t)].
\label{eq:MULTI}
\end{equation}
Note that the first stability multiplier $\lambda_1$ is just that of the
uncoupled 1D map and the coupling affects only the second stability
multiplier $\lambda_2$.

An in-phase orbit is stable only when the moduli of
both multipliers are less than or equal to unity, i.e.,
$-1 \leq {\lambda_i} \leq 1$ for $i=1,2$. A tangent (period-doubling)
bifurcation occurs when each stability multiplier $\lambda_i$ increases
(decreases) through $1$ $(-1)$. Hence the stable region of the in-phase
orbit in the parameter plane is bounded by four bifurcation lines
associated with tangent and period-doubling bifurcations
(i.e., those curves determined by the equations $\lambda_i = \pm 1$
for $i=1,2$).

We now consider the $M$-furcation renormalization transformation $\cal N$,
which is composed of the $M$-times iterating $(T^{(M)})$ and
rescaling $(B)$ operators:
\begin{equation}
{\cal N}(T) \equiv B T^{(M)} B^{-1}.
\label{eq:RON}
\end{equation}
Here the rescaling operator $B$ is:
\begin{equation}
B = \left( \begin{array}{cc}
                    \alpha &  \;\;\;    0\\
                    0      &  \;\;\;    \alpha
                   \end{array}
           \right),
\label{eq:SO}
\end{equation}
because we consider only in-phase orbits.

Applying the renormalization operator ${\cal N}$ to the coupled map
(\ref{eq:CM}) $n$ times, we obtain the $n$-times renormalized map $T_n$ of
the form,
\begin{equation}
 {T_n}:\left\{
       \begin{array}{l}
        x_{t+1} = {F_n}(x_t,y_t) = {f_n}(x_t) + {g_n}(x_t,y_t), \\
        y_{t+1} = {F_n}(y_t,x_t) = {f_n}(y_t) + {g_n}(y_t,x_t).
       \end{array}
     \right.
\label{eq:RTn}
\end{equation}
Here ${f_n}$ and ${g_n}$ are the uncoupled and coupling parts of the $n$-times
renormalized function $F_n$, respectively. They satisfy the following
recurrence equations:
\begin{eqnarray}
&f_{n+1}(x) = &
 \alpha f_n^{(M)} ({\frac {x} {\alpha}}), \label{eq:RUCFn} \\
&g_{n+1}(x,y) =&
             \alpha F_n^{(M)} ({\frac {x} {\alpha}}, {\frac {y} {\alpha}})
          -{\alpha} f_n^{(M)} ({\frac {x} {\alpha}}),
\label{eq:RCFn}
\end{eqnarray}
where
$f_n^{(M)} (x) = f_n (f_n^{(M-1)}(x))$ and
$F_n^{(M)}(x,y) = F_n(F_n^{(M-1)}(x,y), F_n^{(M-1)}(y,x)).$

The recurrence relations (\ref{eq:RUCFn}) and (\ref{eq:RCFn}) define a
renormalization operator $\cal R$ of transforming a pair of functions
$(f,g)$:
\begin{equation}
 \left( \begin{array}{c}
                    {f_{n+1}} \\
                    {g_{n+1}}
                   \end{array}
           \right)
= {\cal R}  \left( \begin{array}{c}
                    {f_n} \\
                    {g_n}
                   \end{array}
           \right).
\label{eq:RORn}
\end{equation}
The renormalization transformation $\cal R$ obviously has a fixed point
$(f^*,g^*)$ with $g^*(x,y)=0$, which satisfies ${\cal R} (f^*,0) = (f^*,0)$.
Here $f^*$ is just the 1D fixed function satisfying
\begin{equation}
f^*(x) = \alpha f^{*(M)}({\frac {x} {\alpha}}),
\label{eq:1DFP}
\end{equation}
where $\alpha = 1 / f^{*(M-1)}(1)$, due to the normalization condition
(\ref{eq:NC}), and it has the form,
\begin{equation}
f^*(x) = 1 + c^*_1 x^2 + c^*_2 x^4 + \cdots,
\end{equation}
where $c^*_i$'s $(i=1,2,\dots)$ are some constants.
The fixed point $(f^*,0)$ governs the critical behavior near the
zero-coupling critical point because the coupling fixed function is
identically zero, i.e., $g^*(x,y)=0$. Here we restrict our attention to this
zero-coupling case.

Consider an infinitesimal perturbation $(h, \varphi)$ to
the zero-coupling fixed point $(f^*,0)$.
We then examine the evolution of a pair of functions $(f^*+h,
\varphi)$ under ${\cal R}$.
Linearizing $\cal R$ at the zero-coupling fixed point, we obtain a linearized
operator $\cal L$ of transforming a pair of perturbations $(h,\phi)$:
\begin{equation}
 \left( \begin{array}{c}
                    {h_{n+1}} \\
                    {\varphi _{n+1}}
                   \end{array}
           \right)
= {\cal L}  \left( \begin{array}{c}
                      {h_n} \\
                     {\varphi_n}
                   \end{array}
              \right)
= \left( \begin{array}{cc}
           {\cal L}_u & \;\;\; 0\\
               0     & \;\;\; {\cal L}_c
           \end{array}
           \right) \;
           \left( \begin{array}{c}
                    {h_n} \\
                    {\varphi_n}
                   \end{array}
              \right),
\label{eq:LOL}
\end{equation}
where
\widetext
\begin{eqnarray}
 h_{n+1}(x) &=&  [{\cal L}_u h_n](x) \\
      &=&  {\alpha} \, \delta f_n^{(M)} ({\frac {x} {\alpha}})
 \equiv {\alpha}\, [f_n^{(M)}({\frac {x} {\alpha}})
         - f_n^{*(M)}({\frac {x} {\alpha}})]_{\rm linear}    \\
     &=& {\alpha} {f^{*}}'(f^{*(M-1)}({\frac {x} {\alpha}})) \,
         \delta f_n^{(M-1)} ({\frac {x} {\alpha}})
         + {\alpha} h_n(f^{*(M-1)}({\frac {x} {\alpha}})), \\
   {\varphi_{n+1}}(x,y) &=& [{\cal L}_c {\varphi_n}](x,y) \\
      &=& {\alpha} \,  {\delta}
      [F_n^{(M)} ({\frac {x} {\alpha}},{\frac {y} {\alpha}} )
        - f_n^{(M)} ({\frac {x} {\alpha}})]
 \equiv {\alpha}\, [F_n^{(M)} ({\frac {x} {\alpha}},{\frac {y} {\alpha}} )
        - f_n^{(M)} ({\frac {x} {\alpha}})]_{\rm linear} \\
     &=& {\alpha} {f^{*}}'(f^{*(M-1)}({\frac {x} {\alpha}})) \,
 {\delta} [F_n^{(M-1)} ({\frac {x} {\alpha}},{\frac {y} {\alpha}} )
        - f_n^{(M-1)} ({\frac {x} {\alpha}})] \nonumber \\
 &&  + {\alpha}  {\varphi_n}(f^{*(M-1)}
  ({\frac {x} {\alpha}}), f^{*(M-1)}({\frac {y} {\alpha}} )).
\end{eqnarray}
\narrowtext
\noindent
Here the variations $\delta f_n^{(M)}
({\frac {x} {\alpha}})$ and ${\delta} [F_n^{(M)} ({\frac {x} {\alpha}},
{\frac {y} {\alpha}})
 - f_n^{(M)} ({\frac {x} {\alpha}})]$ are introduced as the linear terms
(denoted by $[f_n^{(M)}({\frac {x} {\alpha}})
         - f_n^{*(M)}({\frac {x} {\alpha}})]_{\rm linear}$ and
$[F_n^{(M)} ({\frac {x} {\alpha}},{\frac {y} {\alpha}} )
        - f_n^{(M)} ({\frac {x} {\alpha}})]_{\rm linear}$)
in $h$ and $\varphi$ of the deviations of $f_n^{(M)} ({\frac {x} {\alpha}})$
and  $F_n^{(M)} ({\frac {x} {\alpha}},{\frac {y} {\alpha}} )
       - f_n^{(M)} ({\frac {x} {\alpha}})$ from
$f^{*(M)} ({\frac {x} {\alpha}})$ and $0$,
respectively. A pair of perturbations $(h^*,\varphi^*)$ is then called an
eigenperturbation with eigenvalue $\nu$, if it satisfies
\begin{equation}
\nu \, \left( \begin{array}{c}
                    {h^*} \\
                    {\varphi^*}
                   \end{array}
           \right)
= {\cal L}  \left( \begin{array}{c}
                      {h^*} \\
                     {\varphi^*}
                   \end{array}
              \right),
\end{equation}
i.e.,
\begin{eqnarray}
h^*(x) &=& [{\cal L}_u h^*](x), \label{eq:heq}\\
{\varphi}^*(x,y) &=& [{\cal L}_c {\varphi}^* ] (x,y). \label{eq:varphieq}
\end{eqnarray}

The eigenperturbations of the linear operator $\cal L$ can be divided into
two classes. The first class of eigenperturbations are of the form $(h^*,0)$.
Here $h^*(x)$ is an eigenfunction of the linear ``uncoupled operator'' ${\cal
L}_u$
satisfying Eq.~(\ref{eq:heq}), which is just the eigenvalue equation in the
uncoupled 1D case. It has been found in Refs.~\cite{Eckmann,Hao,Chang}
that there exist a unique eigenfunction $h^*(x)$ with (noncoordinate change)
relevant eigenvalue $\delta$, associated with scaling of the nonlinearity
parameter.

The second class of eigenperturbations have the form $(0, {\varphi}^*)$,
where ${\varphi}^*(x)$ is an eigenfunction of the linear ``coupling
operator'' ${\cal L}_c$ satisfying Eq.~(\ref{eq:varphieq}).
However, it is not easy to directly solve the coupling eigenvalue equation
$(\ref{eq:varphieq})$. We therefore introduce a tractable recurrence equation
for a ``reduced coupling eigenfunction '' of $\varphi^*(x,y)$, defined by
\begin{equation}
\left.
\Phi^*(x) \equiv {{\partial {\varphi^*}(x,y)} \over {\partial y}}
\right|_{y=x}.
\label{eq:RCFCT}
\end{equation}
Differentiating Eq.~(\ref{eq:varphieq}) with respect to $y$ and setting
$y=x$, we obtain an eigenvalue equation for a reduced linear
coupling operator $\tilde {{\cal L}_c}$:
\begin{eqnarray}
\nu \, \Phi^*(x) &=& [{\tilde {{\cal L}_c}} {\Phi^*}](x) \label{eq:rlco} \\
&=& \delta F_2^{(M)}({x \over \alpha}) =
[F_2^{(M)}({x \over \alpha})]_{\rm linear} \label{eq:Fvariation}\\
 &=& {f^*}'(f^{*(M-1)}({x \over \alpha}))\,
 \delta F_2^{(M-1)}({x \over \alpha}) \nonumber \\
&& + {f^{*(M-1)}}'({x \over \alpha}) {\Phi^*}(f^{*(M-1)}({x \over \alpha})).
 \label{eq:rceveq}
\end{eqnarray}
Here $F(x,y) = f^*(x) + {\varphi^*}(x,y)$,
$F_2^{(M)}(x)$ is a ``reduced function'' of $F^{(M)}(x,y)$ defined by
$ F_2^{(M)}(x) \equiv {\partial F^{(M)}(x,y)} /
{\partial y} |_{y=x}$,
and the variation $\delta F_2^{(M)}({x \over \alpha})$ is also introduced
as the linear term (denoted by $[F_2^{(M)}({x \over \alpha})]_{\rm linear}$)
in $\Phi^*$ of the deviation of $F_2^{(M)}({x \over \alpha})$ from $0$.

In the case $M=2$, the variation $\delta F_2^{(2)}({x \over \alpha})$
of Eq.~(\ref{eq:Fvariation}) becomes
\begin{equation}
\delta F_2^{(2)}({x \over \alpha}) =
\Phi^*({x \over \alpha}) {f^*}'(f^*({x \over \alpha}))
+ {f^*}'({x \over \alpha}) \Phi^* (f^* ({x \over \alpha})).
\end{equation}
Substituting $\delta F_2^{(2)}({x \over \alpha})$ into
Eq.~(\ref{eq:rceveq}), we have $\delta F_2^{(3)}({x \over \alpha})$
for $M=3$, which consists of three terms,
\begin{eqnarray}
\delta F_2^{(3)}({x \over \alpha})
&=& \Phi^*({x \over \alpha}) {f^*}'(f^*({x \over \alpha}))
  {f^*}'(f^{*(2)}({x \over \alpha})) \nonumber \\
  && + {f^*}'({x \over \alpha}) \Phi^*(f^*({x \over \alpha}))
  {f^*}'(f^{*(2)}({x \over \alpha})) \nonumber \\
  && + {f^*}'({x \over \alpha}) {f^*}'(f^*({x \over \alpha}))
  \Phi^*(f^{*(2)}({x \over \alpha})).
\end{eqnarray}
Repeating this procedure sucessively, we obtain
$\delta F_2^{(M)}({x \over \alpha})$ for a general $M$, composed of
$M$ terms,
\begin{eqnarray}
\delta F_2^{(M)}({x \over \alpha})
 &=& {\sum_{i=0}^{M-1}} {f^{*(i)}}'({x \over \alpha})
 {\Phi^*}(f^{*(i)}({x \over \alpha})) {f^{*(M-i-1)}}'(f^{*(i+1)}
 ({x \over \alpha})) \nonumber \\
 &=& {\Phi^*}({x \over \alpha}) {f^{*(M-1)}}'(f^*({x \over \alpha})) +\cdots
  \nonumber \\
 && + {f^{*(i)}}'({x \over \alpha}) {\Phi^*}(f^{*(i)}({x \over \alpha}))
 {f^{*(M-i-1)}}'(f^{*(i+1)}({x \over \alpha})) + \cdots \nonumber \\
 && + {f^{*(M-1)}}'({x \over \alpha}) {\Phi^*}(f^{*(M-1)}({x \over \alpha})),
\label{eq:F2M}
\end{eqnarray}
where $f^{(0)}(x)=x$.

Using the fact that ${f^*}'(0)=0$, it can be easily shown that when $x=0$,
the reduced coupling eigenvalue equation (\ref{eq:rceveq}) becomes
\begin{equation}
\lambda\, {\Phi^*}(0) = [ {\prod_{i=1}^{M-1}} {f^*}'(f^{*(i)}(0))]\,
     {\Phi^*}(0).
\label{eq:Phi0}
\end{equation}
Differentiating the 1D fixed-point equation (\ref{eq:1DFP}) with respect to
$x$ and then letting $x \rightarrow 0$, we also have
\begin{equation}
{\prod_{i=1}^{M-1}} {f^*}'(f^{*(i)}(0)) =
{\lim_{x \rightarrow 0}} {{f^*}'(x) \over {f^*}'({x \over \alpha})}
=\alpha.
\end{equation}
Then Eq.~(\ref{eq:Phi0}) reduces to
\begin{equation}
 \lambda \, {\Phi^*}(0) = \alpha {\Phi^*}(0).
\end{equation}
There are two cases.
If the coupling eigenfunction $\varphi^*(x,y)$ has a leading linear term,
its reduced coupling eigenfunction $\Phi^*(x)$ becomes nonzero at $x=0$.
For this case $\Phi^*(0) \neq 0$, we have the first CE
\begin{equation}
\nu_1=\alpha.
\label{eq:1stCE}
\end{equation}
The eigenfunction $\Phi_1^*(x)$ with CE $\nu_1$ is of the form,
\begin{equation}
\Phi_1^*(x) = 1 + a_1^* x + a_2^* x^2 +\cdots,
\end{equation}
where $a^*_i$'s $(i=1,2,\dots)$ are some constants.
For the other case $\Phi^*(0)=0$, it is found that ${f^*}'(x)$
is an eigenfunction for the reduced coupling eigenvalue equation
(\ref{eq:rceveq}).
Since Eq.(\ref{eq:F2M}) for the case $\Phi^*(x)={f^*}'(x)$ becomes
\begin{equation}
\delta F_2^{(M)} ({x \over \alpha}) = M {f^{*(M)}}'({x \over \alpha}),
\end{equation}
the reduced coupling eigenvalue equation (\ref{eq:rceveq}) reduces to
\begin{equation}
\nu {f^*}'(x) = M {f^*}'(x).
\end{equation}
We therefore have the second relevant CE
\begin{equation}
\nu_2 = M,
\end{equation}
with reduced coupling eigenfunction $\Phi^*_2(x) = {f^*}'(x)$.
It is also found that there exists an infinite number of additional
(coordinate change) reduced eigenfunctions ${f^*}'(x)\, [f^{*l}(x)-x^l]$
with irrelevant CE's $\alpha^{-l}$ $(l=1,2,\dots)$, which are associated
with coordinate changes. We conjecture that together with the two
(noncoordinate change) relevant CE's $(\nu_1=\alpha,\;\nu_2=M)$,
they give the whole spectrum of the reduced linear coupling operator
$\tilde {{\cal L}_c}$ of Eq.~(\ref{eq:rlco}) and the spectrum is complete.

In order to examine the effect of CE's on the stability multipliers of
periodic orbits in the $M$-furcation sequences, we consider an
infinitesimal coupling perturbation $g(x,y) = \varepsilon \varphi(x,y)$
to a critical map at the zero-coupling critical point, in which case the
two-coupled map has the form,
\begin{equation}
   T:\left\{
       \begin{array}{l}
        x_{t+1} = F(x_t,y_t) = f_{A_\infty^{(M)}}(x_t) + g(x_t,y_t), \\
        y_{t+1} = F(y_t,x_t) = f_{A_\infty^{(M)}}(y_t) + g(y_t,x_t),,
       \end{array}
     \right.
\label{eq:CRM}
\end{equation}
where $A_\infty^{(M)}$ denotes the accumulation value of the parameter $A$
for the $M$-furcation case, and $\varepsilon$ is an infinitesimal
coupling parameter. The map $T$ at $\varepsilon=0$ is just the zero-coupling
critical map consisting of two uncoupled 1D critical maps. It is attracted
to the zero-coupling fixed map consisting of two uncoupled 1D fixed maps
under iterations of the $M$-furcation renormalization transformation
$\cal N$ of Eq.(\ref{eq:RON}).

The reduced coupling function $G(x)$ of
$g(x,y)$ is given by [see Eq.\ (\ref{eq:RCFCT})]
\begin{equation}
G(x)= \varepsilon \Phi(x) \equiv \varepsilon {\displaystyle \left.
{\frac {\partial \varphi(x,y)} {\partial y}}\right|_{y=x}}.
\label{eq:RCFCT2}
\end{equation}
The $n$th image $\Phi_n$ of $\Phi$ under the reduced linear coupling
operator ${\tilde {\cal L}}_c$ of Eq.\ (\ref{eq:rlco}) is of form,
\begin{eqnarray}
\Phi_n (x) &=& [{\tilde{\cal L}}_c^n \Phi](x) \nonumber \\
&\simeq& {\alpha_1} {\nu_1^n} {\Phi^*_1(x)} +
{\alpha_{2}} {\nu_2^n} {f^*}'(x)\;\;{\rm for}\;\;{\rm large}\;\;n,
\end{eqnarray}
because the irrelevant part of $\Phi_n$ becomes negligibly small for
large $n$. Here $\alpha_1$ and $\alpha_2$ are some constants.

The stability multipliers $\lambda_{1,n}$ and $\lambda_{2,n}$ of
the $M^n$-cycle of the map $T$ of Eq.\ (\ref{eq:CRM}) are
the same as those of the fixed point of the $n$-times renormalized map
${\cal N}^n (T)$ \cite{Kim3}, which are given by
\begin{equation}
\lambda_{1,n} = f_n'({\hat x}_n), \;\;\;
\lambda_{2,n} = f_n'({\hat x}_n)- 2 {G_n}({\hat x}_n).
\label{eq:MULTI2}
\end{equation}
Here $f_n$ is the uncoupled part of the $n$th image of
$(f_{A_{\infty}^{(M)}},g)$ under the
renormalization transformation $\cal R$, $G_n(x)$ is the reduced coupling
function of the coupling part $g_n(x,y)$ of the $n$th image, and
${\hat x}_n$ is just the fixed point of $f_n(x)$ [i.e.,
$ {\hat x}_n={f_n}({\hat x}_n)$] and converges to the fixed point $x^*$
of the 1D fixed map $f^*(x)$ as $n \rightarrow \infty$.
In the critical case ($\varepsilon = 0$), $\lambda_{2,n}$ is equal to
$\lambda_{1,n}$ and they converge to the 1D critical stability
multiplier $\lambda^* = {f^*}'(x^*)$.
Since $G_n(x) \simeq  [{\tilde {\cal L}}_c^n G](x)
= \varepsilon \Phi_n(x)$ for
infinitesimally small $\varepsilon$, $\lambda_{2,n}$ has the form
\begin{eqnarray}
{\lambda_{2,n}} & \simeq & {\lambda_{1,n}} -2 \varepsilon \Phi_n \nonumber \\
 &\simeq& {\lambda^*} + {\varepsilon}
\left[ e_1 {\nu_1^n} + e_2 {\nu_2^n} \right]
\;\;{\rm for\;\;large}\;\;n,
\end{eqnarray}
where $e_1 = -2 {\alpha_1} {\Phi^*_1(x^*)}$
and $e_2=-2{\alpha_2} {f^*}'(x^*)$.
Hence the slope $S_n$ of $\lambda_{2,n}$ at the zero-coupling point
($\varepsilon=0$) is
\begin{equation}
S_n \equiv
\left. {\displaystyle {\frac {\partial \lambda_{2,n}}
{\partial \varepsilon} } }\right|_{\varepsilon=0}
\label{eq:slope}
\simeq e_1 {\nu_1^n} + e_2 {\nu_2^n}\;\;{\rm for\;\;large}
\;\;n.
\end{equation}
Here the coefficients $e_1$ and $e_2$ depend on the
initial reduced function $\Phi(x)$, because the constants $\alpha_1$
and $\alpha_2$ are determined only
by $\Phi(x)$. Note that the magnitude of slope $S_n$ increases with $n$,
unless both coefficients $e_1$ and $e_2$ are zero.

We choose monomials $x^l$ $(l=0,1,2,\dots)$ as initial reduced functions
$\Phi(x)$, because any smooth
function $\Phi(x)$ can be represented as a linear combination
of monomials by a Taylor series.
Expressing $\Phi(x) = x^l$ as a linear combination
of eigenfunctions of ${\tilde {\cal L}}_c$, we have
\begin{eqnarray}
\Phi(x) = x^l
&=& {\alpha_1} {\Phi^*_1(x)} +{\alpha_2}
{f^*}'(x) \nonumber \\
&& +{\sum_{l=1}^{\infty}} {\beta_l} {f^*}'(x) [f^{*l}(x)-x^l],
\end{eqnarray}
where $\alpha_1$ is nonzero only for $l=0$, and hence zero for $l \geq 1$,
and all $\beta_l$'s are irrelevant components.
Therefore the slope $S_n$ for large $n$ becomes
\begin{equation}
S_n \simeq \left\{ \begin{array}{l}
           e_1 {\alpha^n} + e_2 M^n \;\;{\rm for}\;\; l=0, \\
                       e_2 M^n\;\;{\rm for}\;\; l \geq 1.
                  \end{array}
           \right.
\label{eq:slope2}
\end{equation}

There are two kinds of coupling. In the case of a linear coupling, in which
the coupling function $\varphi(x,y)$ has a leading linear term, its reduced
coupling function $\Phi(x)$ has a leading constant term. However, for any
other nonlinear-coupling case, in which the coupling function has a
leading nonlinear term, its reduced coupling function contains no
constant term. Hence it follows from Eq.~(\ref{eq:slope2}) that the growth
of $S_n$ for large $n$ is governed by the two relevant CEs $\nu_1=\alpha$
and $\nu_2=M$ for the linear-coupling case ($l=0$), but by only the
second relevant CE $\nu_2=M$ for the other nonlinear-coupling cases
($l \geq 1)$.

\section{Numerical Analysis}
\label{sec:NA}

Taking the trifurcation case with $M=3$ as an example, we numerically
study the scaling behavior associated with coupling in the two coupled 1D
maps (\ref{eq:CRM}) with $f(x)=1-A\,x^2$ and $\varphi(x,y)={1 \over m}
(y^m-x^m)$ $(m=1,2,\dots)$, and confirm the renormalization results.
In this trifurcation case, we follow the $3^n$-cycles up to level $n=9$ and
obtain the slopes of Eq.~(\ref{eq:slope}) at the zero-coupling
critical point $(A_\infty^{(3)}, 0)$
$(A_\infty^{(3)} = 1.786\, 440\, 255\, 563\, 639\,
354\, 534\, 447 \dots)$ when the reduced coupling function $\Phi(x)$ is a
monomial $x^l$ $(l=0,1,\dots)$.

The renormalization result implies that the growth of the slopes $S_n$ is
governed by one CE $\nu_2=3$ for the nonlinear-coupling cases
with $l \geq 1$, i.e., the sequence of $S_n$ obeys a one-term scaling law
asymptotically:
\begin{equation}
   S_n = d_1 r_1^n,
\label{eq:OTSL}
\end{equation}
where $d_1$ is some constant.
We therefore define the growth rate of the slopes as follows:
\begin{equation}
r_{1,n} \equiv {S_{n+1} \over S_n}.
\end{equation}
Then it will converge to a constant $r_1$ as $n \rightarrow \infty$.
As an example consider the case $\Phi(x)=x$.
Figure \ref{figure1} shows three plots of
$\lambda_{2,n}(A_\infty^{(3)},\varepsilon)$
versus $\varepsilon$ for $n=5,6,$ and $7$.
For $\varepsilon=0$, $\lambda_{2,n}$ converges to the 1D critical stability
multiplier $\lambda^*$ $(=1.872\,705\,929 \dots)$
as $n \rightarrow \infty$. However, when $\varepsilon$ is nonzero it diverges
as $n \rightarrow \infty$, i.e., its slope $S_n$ at the zero-coupling critical
point diverges as $n \rightarrow \infty$. The sequence $\{ r_{1,n} \}$ of
the growth rate of $S_n$ is shown in the second column of Table \ref{table1}.
Note that it converges fast to $r_1 = \nu_2 = 3$.
We have also studied two other nonlinear-coupling cases with $l=2,3$ and
found that the sequences of $r_{1,n}$ also converge fast to $r_1 = \nu_2 =3$.

However, in a linear-coupling case with $l=0$, two relevant CE's
$\nu_1=\alpha$ $(=-9.277\,341\dots)$ and $\nu_2=3$ govern the growth of the
slopes $S_n$. We therefore extend the simple one-term scaling law
(\ref{eq:OTSL}) to a two-term scaling law:
\begin{equation}
S_n = d_1 r_1^n + d_2 r_2^n \;\;\;{\rm for\;large\;}n,
\label{eq:TTSL}
\end{equation}
where $d_1$ and $d_2$ are some constants, and $|r_1| > |r_2|$.
This is a kind of multiple scaling law \cite{Mao,Reick}.
The equation (\ref{eq:TTSL}) gives
\begin{equation}
S_{n+2} = q_1 S_{n+1}  - q_2 S_n,
\label{eq:TTRE}
\end{equation}
where $q_1 = r_1 + r_2$ and $q_2 = r_1 r_2$.
Then $r_1$ and $r_2$ are solutions of the following quadratic
equation,
\begin{equation}
r^2 - q_1 r + q_2 = 0.
\label{eq:EVEr}
\end{equation}
To evaluate $r_1$ and $r_2$, we first obtain $q_1$ and $q_2$
from $S_n$'s using Eq.\ (\ref{eq:TTRE}):
\begin{equation}
q_1 = {\frac { S_{n+1} S_n - S_{n+2} S_{n-1} } {S_n^2-S_{n+1} S_{n-1}} },\;\;
q_2 = {\frac { S_{n+1}^2 - S_n S_{n+2} } {S_n^2-S_{n+1} S_{n-1}} }.
\label{eq:t12}
\end{equation}
Note that Eqs.\ (\ref{eq:TTSL})-(\ref{eq:t12}) are valid for large $n$.
In fact, the values of $q_i$'s and $r_i$'s $(i=1,2)$ depend on the level $n$.
Thus we denote the values of $q_i$'s in Eq.\ (\ref{eq:t12})
explicitly by $q_{i,n-1}$'s, and the values of $r_i$'s obtained
from Eq.\ (\ref{eq:EVEr}) are also denoted by $r_{i,n-1}$'s.
Then each of them converges to a constant as $n \rightarrow \infty$:
\begin{equation}
\lim_{n \rightarrow \infty} q_{i,n} = q_i,\;\;
\lim_{n \rightarrow \infty} r_{i,n} = r_i,\;\;i=1,2.
\end{equation}

When $\Phi(x)=1$, plots of $\lambda_{2,n}(A_\infty^{(3)},\varepsilon)$
versus $\varepsilon$ for $n=2,3,$ and $4$ are shown in Fig.~\ref{figure2}.
The slopes $S_n$ at $(A_\infty^{(3)},0)$ obeys well the
two-term scaling law (\ref{eq:TTSL}).
Sequences $\{ r_{1,n} \}$ and $\{ r_{2,n} \}$ are shown in the third and
fourth columns of Table \ref{table1}.
Note that they converge fast to $r_1 = \nu_1 = \alpha$
and $r_2 = \nu_2 = 3$, respectively.

\section{summary}
\label{sec:summary}
The scaling behavior of $M$-furcations is studied in two symmetrically
coupled 1D maps. Using a renormalization method, the dependence of the
scaling behavior on $M$ is particularly investigated in the zero-coupling
case. It is found that the zero-coupling fixed map of the $M$-furcation
renormalization operator has three relevant eigenvalues $\delta$,
$\alpha$, and $M$. As in the case of 1D maps, the eigenvalue $\delta$
governs the scaling behavior of the nonlinearity parameter,
irrespectively of coupling.
However, the scaling behavior associated with coupling
depends on the nature of coupling. In a linear-coupling case,
it is governed by two CE's $\alpha$ and $M$, whereas it is governed by
only one CE $M$ in the case of a nonlinear-coupling case. Taking the
trifurcation case as an example, we also study the coupling effect on
the second stability multipliers of $3^n$-cycles by a direct numerical
method and confirm the renormalization results.

\acknowledgments
This work was supported by the Basic Science Research Institute Program,
Ministry of Education, Korea, Project No. BSRI-94-2401.

%
%

\begin{table}
\caption{In a nonlinear-coupling case $\Phi(x)=x$, a sequence
$\{ r_{1,n} \}$ for a one-term scaling law is shown in the second column,
and in the linear coupling case $\Phi(x)=1$, two sequences $\{ r_{1,n} \}$
and $\{ r_{2,n} \}$ for a two-term scaling law are shown in the third and
fourth columns, respectively.}
\begin{tabular}{cccc}
& \multicolumn{1}{c}{$\Phi(x)=x$} & \multicolumn{2}{c}{$\Phi(x)=1$} \\
$n$  & $r_{1,n}$ & $r_{1,n}$ &  $r_{2,n}$ \\
\tableline
1 &  2.997\,929\,154  &  -9.276\,543\,16   & 2.005\,8 \\
2 &  3.000\,141\,141  &  -9.277\,415\,78   & 2.692\,5 \\
3 &  2.999\,990\,417  &  -9.277\,335\,54   & 2.927\,8 \\
4 &  3.000\,000\,651  &  -9.277\,341\,50   & 2.984\,5 \\
5 &  2.999\,999\,956  &  -9.277\,341\,09   & 2.996\,7 \\
6 &  3.000\,000\,003  &  -9.277\,341\,12   & 2.999\,3 \\
7 &  3.000\,000\,000  &  & \\
8 &  3.000\,000\,000  &  & \\
\end{tabular}
\label{table1}
\end{table}
%
%

\begin{figure}
\caption{Plots of the second stability multipliers
$\lambda_{2,n} (A_\infty^{(3)},\varepsilon)$ versus $\varepsilon$
        for $n=5,6,7$ in a nonlinear-coupling case $\Phi(x)=x$.}
\label{figure1}
\end{figure}
\begin{figure}
\caption{Plots of the second stability multipliers
$\lambda_{2,n} (A_\infty^{(3)},\varepsilon)$ versus $\varepsilon$
        for $n=2,3,4$ in the linear-coupling case $\Phi(x)=1$.}
\label{figure2}
\end{figure}
%
%

\end{document}